# Defects encode high-dimensional topological information

Yunqi Zhang[1,†], Fengjun Li[2,†], Runchen Zhang[1,†], Zi-Lan Deng[2,†], Liangyu Deng[1], Zhikai Zhou[2], Ruofu Liu[1], Zimo Zhao[1], Yifei Ma[1], Yuanzhe Xu[1], Zixuan Wang[3], Yixuan Zhao[3], Jize Yan[3], Honghui He[4], Xiangping Li[2,5,*], and Chao He[1,*]

[1]*Department of Engineering Science, University of Oxford, Parks Road, Oxford, OX1 3PJ, UK*
[2]*Guangdong Provincial Key Laboratory of Micro-nano Optical Precision Fabrication Technologies and Applications, Institute of Photonics Technology, Jinan University, Guangzhou 510632, China*
[3]*School of Electronics and Computer Science, University of Southampton, Southampton, SO17 1BJ, UK*
[4]*Guangdong Research Center of Polarization Imaging and Measurement Engineering Technology, Tsinghua Shenzhen International Graduate School, Tsinghua University, Shenzhen 518055, China*
[5]*Nanhu Laser Laboratory, Changsha 410073, China*
[†]*These authors contributed equally to this work*
[*]*Corresponding authors: xiangpingli@jnu.edu.cn, chao.he@eng.ox.ac.uk*

**Abstract**

**In polarization fields, Stokes skyrmions are continuous vectorial textures that encode integer-valued topological invariants across real space, enabling robust optical information encoding under complex perturbations. This topological resilience, however, fails when singular points occur where the Stokes vector has no unique limiting value, placing a fundamental constraint on skyrmion-based information manipulation. Here, we show, paradoxically, that the very defects that destroy conventional resilience can become the carriers of topological information. We introduce the resulting structures as Stokes defect skyrmions, in which singular Stokes responses constitute measurable topological degrees of freedom with theoretically minimal size. We design and realize one class of them using all-dielectric metasurfaces that combine arbitrarily controlled distinguished fast-axis singularities with customized retardance profiles. The resulting fields are then described by high-dimensional integer-valued topological tuples, providing theoretically unbounded information capacity at the nanoscale. As a proof-of-concept demonstration, selected tuple components are mapped to represent predefined alphabetic symbols, realizing controlled high-dimensional information representation within a single optical field. Our results establish Stokes defects as functional units for higher-dimensional topological encoding, expanding the role of defects from failure points to engineerable carriers of optical information.**

Polarization provides an intrinsically vectorial degree of freedom for structuring optical information[1–5], allowing information to be encoded in a broader optical state space than scalar field variables such as intensity or phase[6–8]. This potential has been advanced by Stokes skyrmions[9–14], which endow polarization fields with integer-valued topological invariants that can remain stable under perturbations[15–19], making perturbation-resilient optical information encoding a concrete possibility[20–22]. Yet, the resilience offered by Stokes skyrmions rests on a stringent condition: the underlying Stokes field must be a continuous and compactifiable mapping onto the Poincaré sphere[15].

This assumption fails when singular points occur within the Stokes field. At such locations, the limiting normalized Stokes vector depends on the direction from which the point is approached, so no unique value can be assigned; formally, $\lim_{\boldsymbol{r}\to\boldsymbol{r}_0} \boldsymbol{S}(\boldsymbol{r})$ does not exist. These singularities are termed Stokes defects. Recent studies have shown that such defects can cause the skyrmion number to lose its integer-invariant meaning[23,24], leading to unstable or incorrectly encoded information.

Here, rather than treating Stokes defects as failure points of skyrmion topology, we use them as localized topological units that enrich the Stokes field with theoretically minimal spatial cost. Since each defect is confined to an isolated point, additional defects can introduce extra topological degrees of freedom without requiring a proportional increase in optical area (see Methods 2&3). We term Stokes fields that incorporate such defects into skyrmionic textures Stokes defect skyrmions. To describe these textures, we use the generalized skyrmion framework[25,26], assigning defect-bearing Stokes fields an integer-valued tuple, $(a_1, a_2 \cdots a_k) \in \mathbb{Z}^k$. This tuple is constructed by using defect-induced boundary trajectories, together with the aperture boundary, to partition the Poincaré sphere into connected regions. The tuple components are regional invariants jointly determined by the trajectory geometry and the Stokes-field configuration. Each component can be separately manipulated and extracted within a single optical field (see Supplementary Note 6 and Ref.[25] for details).

To engineer Stokes defect skyrmions on demand, we exploit the singular anisotropic response of spatially structured matter, using all-dielectric $TiO_2$ metasurfaces throughout this work. In such systems, the azimuthal winding required for Stokes textures which determines the $a_k$ can be generated by a spatially varying fast-axis orientation, which by design contain singularities in the device plane (see Methods Fig. 4b). Conventional designs for continuous Stokes fields suppress their effect by choosing a degenerate retardance, such as an integer multiple of $2\pi$, for which the optical response becomes independent of the undefined fast-axis orientation and the output Stokes field remains well defined[9,27–32]. By deliberately relaxing this constraint, we create direction-dependent local optical responses and prescribe the number, position and winding order of fast-axis singularities, thereby controlling the resulting topological tuple encoded in the matter (see Methods 1, Supplementary Notes 4 and 5 for metasurface design and fabrication details).

We realize this strategy experimentally using designed Mueller matrix responses to generate prescribed defect configurations. Across a hierarchy of designs, including fields with up to four Stokes defects, the measured tuple-valued invariants agree well with theoretical predictions, and a level of error resilience is demonstrated. Finally, we demonstrate a proof-of-concept topological information encoding by mapping selected tuple components onto predefined alphabetic symbols, enabling a single Stokes defect skyrmion (four defects) to represent a multi-symbol information state. We note the distinct states increases exponentially with the number of defects, while the corresponding encoding space expands to $\mathbb{Z}^{N+1}$ for $N$ defects. These results establish engineered Stokes defects as potential ultra-high-information-density topological carriers (see Methods 3), enlarging the information capacity without requiring a proportional increase in optical footprint.

## Results

At the heart of our method is the design and characterization of Stokes defects themselves, as well as their generation through engineered matter field construction. The concept of how a designed Stokes defect skyrmion can encode tuple-valued topological invariants is shown in Fig. 1a. For uniformly polarized light (Fig. 1a(i)), the field covers only a single point on the 2-sphere $S^2$, i.e. the Poincaré sphere; the corresponding topology is therefore trivial. In Fig. 1a(ii), a conventional Néel-type Stokes skyrmion covers the whole $S^2$, leading to the conventional skyrmion number[10,33] of $N_{sk}$=1. Extending this idea, Fig. 1a(iii) shows a designed defect whose coloured boundary trajectory has an azimuthal winding number of one and partitions the Poincaré sphere into two connected regions, yielding the tuple invariant $N_{G_sk} = (0, 1)$. The detailed mechanism with the generalized skyrmion theory can be found in Supplementary Note 6.

We would like to highlight that this tuple invariant not only assigns multiple numbers to the defects, but itself also features strong error resilience. In practical settings, disorder may substantially reshape the related Stokes texture and continuously deform the corresponding trajectory on $S^2$ (see examples in Fig. 1b). However, those deformations do not alter the encoded information, provided that the boundary trajectory does not cross a topological transition that changes the sphere partition[25]. The protected topological quantity can therefore be explained in the sense that it is not determined by the exact shape of the Stokes field boundary, but by the defect-induced partition. Such a relationship between geometrical deformation and topological change underpins the experimental robustness and reconstruction results below.

To create these defects in practice, we formulate the design from the matter side: the distribution of fast-axis orientation singularities determines the defect number, position and winding order, while the local retardance controls how these axis geometries are converted into boundary trajectories on $S^2$ (see Methods 1 and Supplementary Note 4 for design process details). We realize this strategy experimentally using all dielectric $TiO_2$ metasurfaces. Representative scanning electron microscopy (SEM) images are presented in Fig. 1c, alongside the demonstration of a Stokes defect skyrmion comprising three defects.

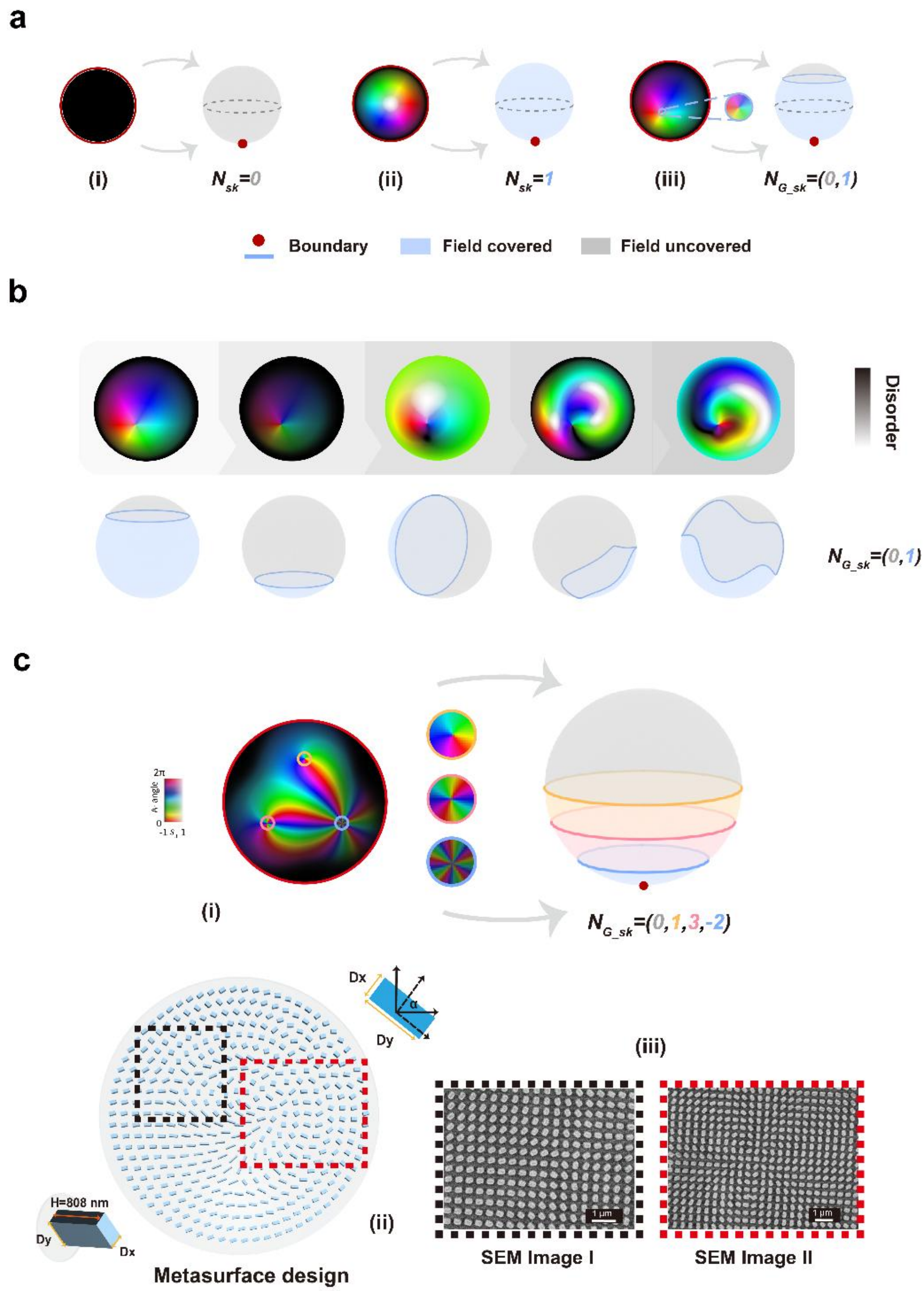

Fig. 1: **Concept and implementation of Stokes defect skyrmions. a.** Construction of a Stokes defect skyrmion. The red point on the Poincaré sphere represents the image of the aperture boundary, whereas the blue curve represents the boundary trajectory induced by a Stokes defect. Light blue and gray denote regions covered and uncovered by the Stokes field, respectively (Stokes vector field is visualised using a hue–lightness colour scheme, in which the hue encodes the azimuthal angle and the lightness encodes $s_3$). (i) A trivial left circularly polarised (LCP) field collapses to a single point and gives $N_{sk} = 0$, whereas (ii) a conventional Néel-type skyrmion covers a complete spherical region, giving $N_{sk} = 1$. (iii) For a Stokes defect skyrmion, the defect-induced trajectory partitions the sphere into distinct connected regions, which are assigned separate integer invariants to form $N_{G_sk} = (0,1)$. **b**. Topological resilience under increasing disorder. Although the Stokes texture and the boundary trajectory geometry deform, the partition of the Poincaré sphere remains topologically unchanged. The recovered tuple therefore remains unchanged throughout the perturbation regime (see Supplementary Note 7 for details). **c**. Multiple defects and metasurface implementation of Stokes defects. (i) The example contains three engineered defects within a single optical field, producing additional connected regions on the Poincaré sphere. Note that each value of the topological tuple is obtained by cumulative addition, rather than from the winding number of the individual defect: this leads to the blue-

circled singularity, which has a winding number of 5, ending up with a tuple element value of $-2$ (see Supplementary Note 6 for more details). (ii) The target response is encoded by the nanopillar parameters $D_x$, $D_y$, the orientation[34] $\alpha$ and the height of the pillar $H$(fixed at 808 nm). Representative SEM images of the two marked regions show the fabricated nanopillar arrangements, scale bar is 1 $\mu$m.

We then first experimentally validated this defect-engineering strategy through a single designed Stokes defect. The procedure for realizing and characterizing the Stokes defect metasurface is as follows. First, we designed and fabricated the corresponding metasurface using the developed language for Stokes defects. Second, both the Mueller response of the metasurface and the output Stokes field were measured by using Stokes–Mueller polarimetry[35–38] (see Fig. 2a, Supplementary Notes 2&3 for experimental setup and details). Third, to examine the defect properly, we selected a small contour enclosing the defect core and mapped the simulated and measured Stokes vectors along this loop onto the Poincaré sphere. The results are shown in Fig. 2b, where the measured metasurface response closely reproduces the designed response. The fast-axis orientation is well matched with the prescribed distribution, while the retardance exhibits modest deviations, primarily arising from metasurface fabrication-related errors. On the beam side, the output Stokes components $S_1$, $S_2$ and $S_3$, together with the resulting trajectory on the Poincaré sphere, reproduce the overall features of their simulated counterparts (see Fig. 2c & d for quantitative analysis).

Consequently, the topological tuple was calculated from the measured field, yielding $N_{G_sk} = (0,1.00)$, which agrees well with the designed theoretical value. We note that the measured trajectory on the Poincaré sphere is visibly deformed from the theoretical one due to fabrication and measurement errors. Together with the correctly obtained topological tuple, this validates the robustness of Stokes defect skyrmions against experimental imperfections.

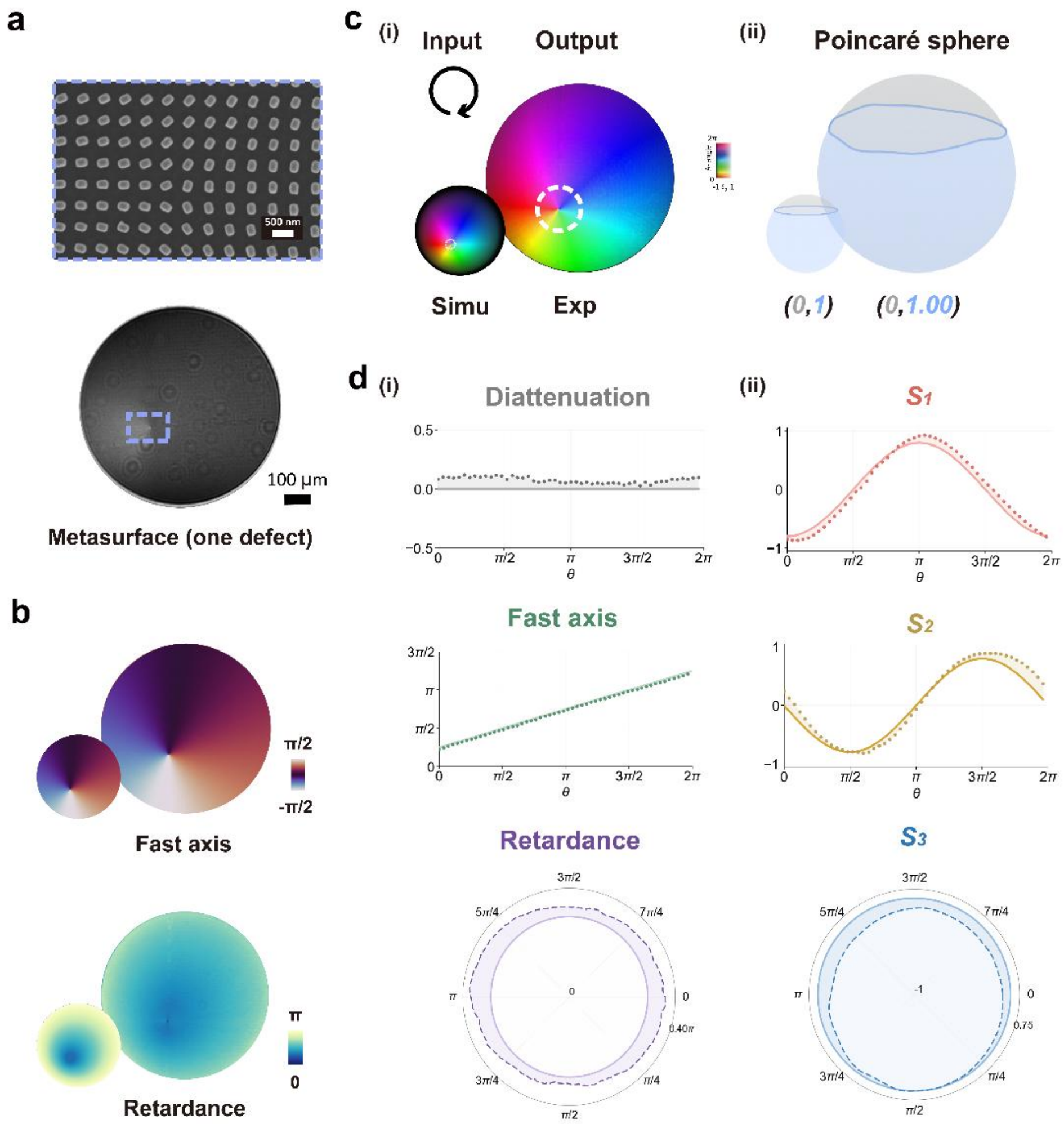

Fig. 2: **Experimental realization and characterization of a single Stokes defect skyrmion. a.** SEM image (top) corresponding to the blue-highlighted region in the measured intensity image (bottom) of the metasurface containing one designed singularity. Scale bars are 500 nm and 100 $\mu$m respectively. **b.** Simulated and measured fast-axis orientation and retardance profiles across the aperture. **c.** (i) Simulated and measured Stokes fields generated under circular polarized illumination. The white dashed circle marks the contour surrounding the Stokes defect used to extract the defect trajectory. (ii) The corresponding trajectories on the Poincaré sphere preserve the same partition, yielding the prescribed tuple $N_{G_sk} = (0,1.00)$. **d.** (i) Diattenuation, fast-axis orientation and retardance sampled along the circular contour indicated in **c** as functions of the azimuthal coordinate $\theta$. Solid curves show the simulated results, dashed curves show the experimental results, and the shaded regions indicate their pointwise errors. (ii) Contour-resolved comparison of the Stokes components $S_1$, $S_2$, and $S_3$ around the defect, using the same line and shading conventions as in **d(i).**

We then demonstrate the multi-defect scenarios, aiming to show the dimensionality and density of encoded information afforded by the defect architecture. Here, we first adopt a one-defect-one-dimension strategy: we note that each additional engineered defect introduces one additional tuple component (represented on the Poincaré sphere by an additional non-intersecting trajectory), expanding the encoding space from $\mathbb{Z}^2$ with one defect to $\mathbb{Z}^{N+1}$ with $N$ defects (Fig. 3a) (see Methods 2 and 3 for other defect structure designs and encoding densities). We demonstrate this multidimensional control by encoding the four-character string **OXJN**, representing the University of Oxford and Jinan University, into a single optical field.

The workflow is as follows. First, we mapped the 26 letters of the alphabet onto a predefined set of integer labels selected for reliable metasurface realization and optical readout (Fig. 3b). Second, we designed and fabricated a single metasurface containing four prescribed fast-axis singularities together with the corresponding retardance distribution. Third, using the same methods as in the previous single-field characterization, we reconstructed the output Stokes field, extracted the corresponding defect configurations (Fig. 3c), and calculated the topological tuple (Fig. 3d).

As shown, the topological tuple is obtained as $N_{G_sk} = (0, -1.02, 1.00, -4.05, -2.04)$, well matched with the designed values. This result validates two key capabilities of the Stokes defect skyrmion framework for topological encoding: (1) each component of the topological tuple can be independently controlled and encode information through coordinated tuning of the defect parameters, and (2) the dimensionality of the topological tuple can be increased by adding engineered defects, with the total information density increasing correspondingly and the tuple space extending towards $\mathbb{Z}^{N+1}$. Together, these results show that both the encoded information and its dimensionality can be directly controlled through the engineered defect configuration.

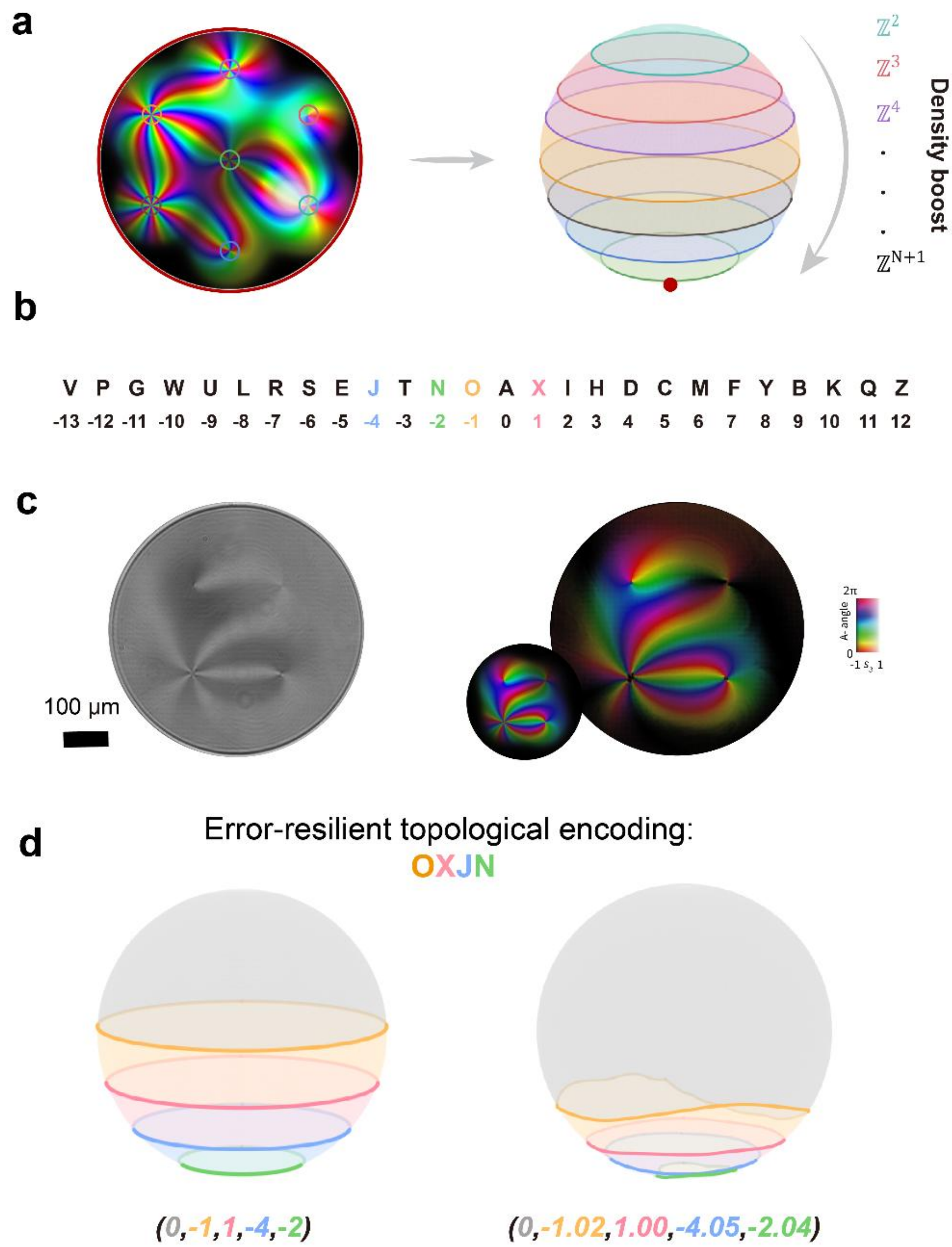

Fig. 3: **High-dimensional topological encoding with multiple Stokes defects. a**. Exponential growth of topological state enabled by the linear increase in engineered defects. For the non-intersecting trajectories, each additional engineered defect introduces one further connected region and increases the tuple dimension by one, providing a direct route from $\mathbb{Z}^2$ to progressively higher-dimensional integer spaces $\mathbb{Z}^{N+1}$, where $N$ is the number of defects. **b**. Alphabetic encoding scheme used for the proof-of-concept demonstration. The 26 letters are mapped onto distinct integers. **c**. Experimental realization of a multi-defect Stokes field. A single metasurface encodes four fast-axis orientation singularities and generates four Stokes defects within one Stokes defect skyrmion, as confirmed by the reconstructed Stokes texture. Scale bar for the intensity profile is 100 $\mu$m. **d**. Simulated and measured boundary trajectories of the introduced boundaries on the Poincaré sphere. As the number of defects increases, fabrication and measurement errors have a greater effect on the more complex field, leading to larger distortions in the measured trajectories. Nevertheless, the boundary-induced partition remains unchanged, and the tuple is recovered as $N_{G_sk} = (0, -1.02, 1.00, -4.05, -2.04)$, with all components retaining their prescribed integer values and yielding the encoded string **OXJN**.

## Discussion

This work establishes Stokes defect skyrmions as engineered topological textures built around singular points where the Stokes vector has no well-defined limiting value. By spatially patterning Mueller matrix responses onto all-dielectric $TiO_2$ metasurfaces with direction-dependent limits, we show that such defects can be deliberately embedded in Stokes fields and used as functional units for high-dimensional information manipulation. In this way, defects that would normally disrupt skyrmion quantization are instead turned into useful topological degrees of freedom. Our demonstration further suggests that these defect-based invariants can remain topologically robust during transmission through distorting optical media (see Supplementary Note 7), motivating further investigation of this mechanism in more complex fields, such as those containing multiple defects. We note that the present encoding experiments are proof-of-concept validations. Further practical implementation will require more efficient symbol mapping and more robust extraction of defect boundaries and topological invariants under noise[39]. These are, however, engineering challenges rather than fundamental limits of the framework: data-driven[40–43] and physics-informed inference methods[44–46] may help recover topological invariants rapidly and robustly from imperfect measurements.

Several directions naturally follow from this topological encoding method. First, the configurations demonstrated here are restricted to non-intersecting trajectories under a uniform boundary condition, providing a controlled setting in which the invariant tuple can be defined and measured unambiguously. More general winding (see Methods 2 and 3), intersection and self-intersection geometries can partition the Poincaré sphere into more regions using the same number of defects, revealing that the information density depends not only on the number of defects but also on the geometry of their trajectories. For instance, different trajectory geometries (such as intersecting trajectories) can increase the invariant space from $\mathbb{Z}^{N+1}$ to $\mathbb{Z}^{N^2-N+2}$, even with the same number of defects; a self-intersecting trajectory can generate multiple tuple components from a single defect and, in principle, is already able to increase the tuple space to $\mathbb{Z}^{\infty}$ (see Methods 3 and Supplementary Note 1). This opens a much broader design space for advanced information encoding using Stokes defect skyrmions.

Second, although the demonstrations in this paper use $TiO_2$ metasurfaces, the developed framework is not tied to this platform: i.e., any polarization-engineering system capable of controlling a spatially varying Mueller-matrix response could realize such defect-bearing Stokes fields. In this way, possible implementations such as passive laser-written birefringent structures[28,47–49] and reconfigurable liquid-crystal systems[50–54] can be customized for generation in an application-guided manner. Furthermore, Stokes defects need not arise only from fast-axis singularities in linear retarders; other types of polarisation elements can also play such roles. For example, singularities arising from spatially varying elliptical-retarder responses, or from spatially structured diattenuation[26,31], could provide alternative physical routes. These different implementations offer intriguing scope for further exploration.

Third, the concept of defect-shaped structures can extend beyond structured light to other physical fields, including liquid-crystal director fields[55–57], magnetic spin textures[58,59], and optical-axis fields in anisotropic media[21,26,31,60]. Even when their parameter spaces are not naturally spherical, suitable and physically meaningful mappings could project these distributions onto an effective spherical space[60]. In this way, the defect-enabled topological framework could be extended to different physical systems, offering possible routes towards high-density information storage in permanent or rewritable media, from laser-written glass to dynamically addressable platforms.

Fourth, beyond these possible extensions, we would like to specifically note that a central feature of Stokes defects is their particularly rich information structure: their vectorial nature supports complex configurations with multiple topological degrees of freedom. This capability comes from the defects themselves, while the generalized skyrmion formalism used here just provides one way to assign integer values to them. We note that **other topological constructions** (and related numbers; see Supplementary Note 8 for other experimental implementations) that associate invariants with singularities could serve the same purpose.

Overall, our work provides a defect-based information encoding strategy in which the information capacity is directly controlled by the defect configuration. This offers a simple route towards high-dimensional, ultra-high-density topological information processing on compact platforms.

## Methods

### Metasurface design and meta-atom library mapping strategy

In this work, a set of $TiO_2$ metasurfaces was designed to generate the Stokes defect skyrmions by controlling the polarization state of the transmitted light. The target matter field is parameterized by spatially varying distributions of fast-axis orientation $\alpha(x,y)$ and retardance $\Delta\varphi(x,y)$, which together specify the required anisotropic (Mueller or Jones) response. Each meta-atom is therefore treated as a high-transmission, nondepolarizing retarder element with minimal diattenuation. The $\alpha(x,y)$ and $\Delta\varphi(x,y)$, profiles are then discretized onto the metasurface configuration: the fast-axis orientation is encoded by the in-plane rotation of each $TiO_2$ nanopillar (Fig. 1c), whereas the retardance is realized through the relative phase delay between its two orthogonal linear eigenpolarizations, controlled by the nanopillar dimensions. The Jones matrix can be expressed as[34]

$$J(x,y) = e^{i\varphi(x,y)} R[\alpha(x,y)] \begin{bmatrix} A_x(x,y) e^{i\frac{\Delta\varphi(x,y)}{2}} & 0 \\ 0 & A_y(x,y) e^{-i\frac{\Delta\varphi(x,y)}{2}} \end{bmatrix} R[-\alpha(x,y)] \tag{1}$$

where $\varphi(x,y)$ is the overall phase delay, $R$ is the $SO(2)$ rotation matrix, $A_x$ and $A_y$ are the modulated amplitudes. (see Supplementary Note 5) To determine the metasurface design parameters, we employ a meta-atom library mapping strategy. The selected meta-atoms are required to satisfy two sets of constraints. First, they should exhibit high and nearly polarization-independent transmission, $|A_x - A_y| \approx 0$, while maintaining a spatially uniform[61–63] $\varphi(x,y) = \phi_0$. Second, the relative phase delay of each meta-atom must reproduce the prescribed retardance profile $\Delta\varphi(x,y)$. However, directly imposing a strict amplitude-balance constraint on the meta-atom library would substantially reduce the geometric parameter space and could leave parts of the required retardance range unmapped.

To satisfy all these requirements, we develop a robust method for identifying the geometric parameters without imposing additional constraints on the parameter range. First, the low-efficiency region is removed from the meta-atom library, and only meta-atoms with efficiencies exceeding $80\%$ are retained. An overall phase $\varphi$ is then selected and combined with $\Delta\varphi$ to calculate the required phase retardations $\varphi_x$ and $\varphi_y$ for the desired meta-atoms. To determine the optimal geometric parameters, we define a root-mean-square-error ($RMSE$) function for rapid mapping. The $RMSE$ function is given by

$$RMSE = \sqrt{(\varphi_x - \varphi_x^{\mathrm{lib}})^2 + (\varphi_y - \varphi_y^{\mathrm{lib}})^2} \tag{2}$$

where $\varphi_x^{\mathrm{lib}}$ and $\varphi_y^{\mathrm{lib}}$ are the phase retardations obtained from the meta-atom library. The desired parameters correspond to the minimum value of $RMSE$. In general, the meta-atoms obtained from an initial mapping do not provide the required amplitude balance. To address this issue, we optimize the overall phase $\varphi$ and repeat the mapping process to identify the optimal parameters.

Applying this optimization method to a Stokes defect skyrmion containing two defects, we obtain the metasurface design summarized in Fig. 4. A total of 104 distinct meta-atom geometries were selected to construct the metasurface (Figs. 4d, 4e and 4f). The overall phase distribution $\varphi(D_x, D_y)$ of the meta-atom library is shown in Fig. 4d. By optimizing $\varphi(x,y)$, the selected meta-atoms are mapped onto the corresponding iso-phase contour. For the optimized value $\phi_0 = 0$, the selected meta-atoms fully cover the prescribed retardance range, while the amplitude difference $|A_x - A_y|$ remains close to zero. This confirms that the selected meta-atoms exhibit a low-diattenuation retarder response. Fig. 4g also shows the $RMSE$ distribution of the selected meta-atoms, with values reaching the order of $10^{-2}$. The transmittance distributions of $T_x$ and $T_y$ are also shown in Fig. 4h & 4i. Both remain above $85\%$ for all selected meta-atoms, and more than half of the selected geometries exhibit transmittances above $90\%$.

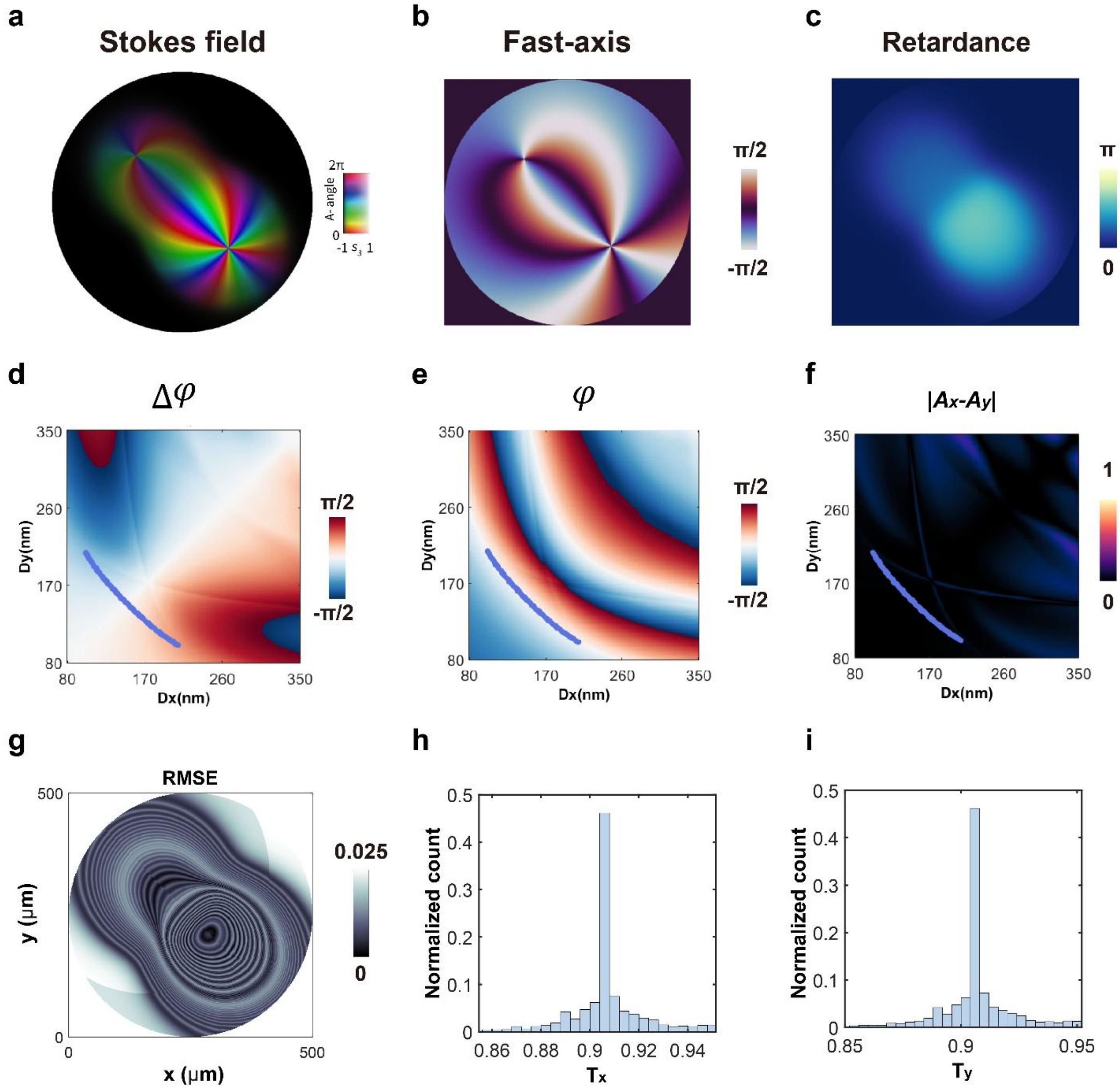

Fig. 4: **Meta-atom selection strategy. a.** Simulated Stokes field of a two-defect Stokes defect skyrmion (see Supplementary Note 5 for experimental results) **b.** simulated fast-axis orientation of the matter field. **c.** simulated retardance profile of the matter field **d.** Distribution of the phase retardance $\Delta\varphi = \varphi_x - \varphi_y$ in the $(D_x, D_y)$ parameter space. Blue markers indicate the 104 selected meta-atom types. **e.** Distribution of the optimized overall phase $\varphi$ in the same parameter space. The overall phase is fixed at zero in the final design. **f.** Amplitude imbalance $|A_x - A_y|$ across the parameter space. The near-zero values at all selected meta-atom locations (blue markers) confirm that each meta-atom functions as an effective retarder. **g.** Spatial distribution of the RMSE across the metasurface aperture. The error remains as low as $10^{-2}$ demonstrating high fidelity of the mapping procedure. **h.** and **i.** Transmittance histograms all selected meta-atoms. All exceed $85\%$ efficiency, with more than half surpassing $90\%$.

## Construction of topological tuple from engineered boundary trajectories

This method mainly visualizes how the encoded topological information can be decoded from the boundaries (and defects) information of a designed Stokes field through illustrations on the Poincaré sphere. A Stokes defect skyrmion can be characterized by the trajectories associated with its aperture boundary and the defects (in effect extra boundaries). Once the experimental Stokes data are obtained, and mapped onto the Poincaré sphere, these trajectories form a curve set that partitions the sphere into connected regions. The number of connected regions determines the dimension of the topological tuple, whereas the full Stokes field determines the integer assigned to each region (see Supplementary Note 6 for details). The collection of these regional invariants forms the topological tuple $N_{G_sk}$.

Figure 5 presents several representative boundary geometries and their corresponding topological tuples. Fig. 5a shows a degenerate case of the *L-line*[64,65], for which $s_3 = 0$ throughout the transverse plane and the complete Stokes field maps onto the equator of the Poincaré sphere. The aperture boundary and the defect-induced boundary therefore collapse onto the same equatorial curve. Although this trajectory formally separates the sphere into two hemispheres, the Stokes field remains confined to the boundary itself and does not cover either region. Both regional invariants consequently vanish, giving $N_{G_sk} = (0,0)$. We note a non-degenerate Stokes defect skyrmion requires the image of the Stokes field to have nonempty coverage within at least one connected region of the boundary-partitioned Poincaré sphere.

In Fig. 5b, the boundary traces a closed curve $\mu(t)$ below the equator, while a single defect-generated trajectory $\gamma(t)$ runs along the upper hemisphere. Together, these two curves partition $S^2$ into three connected regions, and the whole Stokes field yields the tuple $N_{G_sk} = (3,-2,0)$. In Fig. 5c, intersections between existing boundary curves refine the partition of the Poincaré sphere. Each crossing becomes a vertex of the boundary graph, and the regions enclosed between the intersecting segments acquire their own integer invariants. In the example shown, the $\mu(t)$ and $\gamma(t)$ intersect at two points, subdividing what would otherwise be two regions into four, and the tuple becomes $N_{G_sk} = (-1,2,1,-2)$. Fig. 5d shows a self-intersecting case. Here, a defect-associated trajectory crosses itself and generates several connected regions from a single curve, producing $N_{G_sk} = (4,-2,0,2)$.

Taken together, these configurations show that the dimension of the topological tuple is governed by the full boundary geometry mapped onto the Poincaré sphere, whereas the value of each component is determined by how the Stokes field covers the corresponding region. Thus, an important implication is that fields containing the same number of physical defects can support different tuple structures when their trajectories wind, intersect, or self-intersect differently.

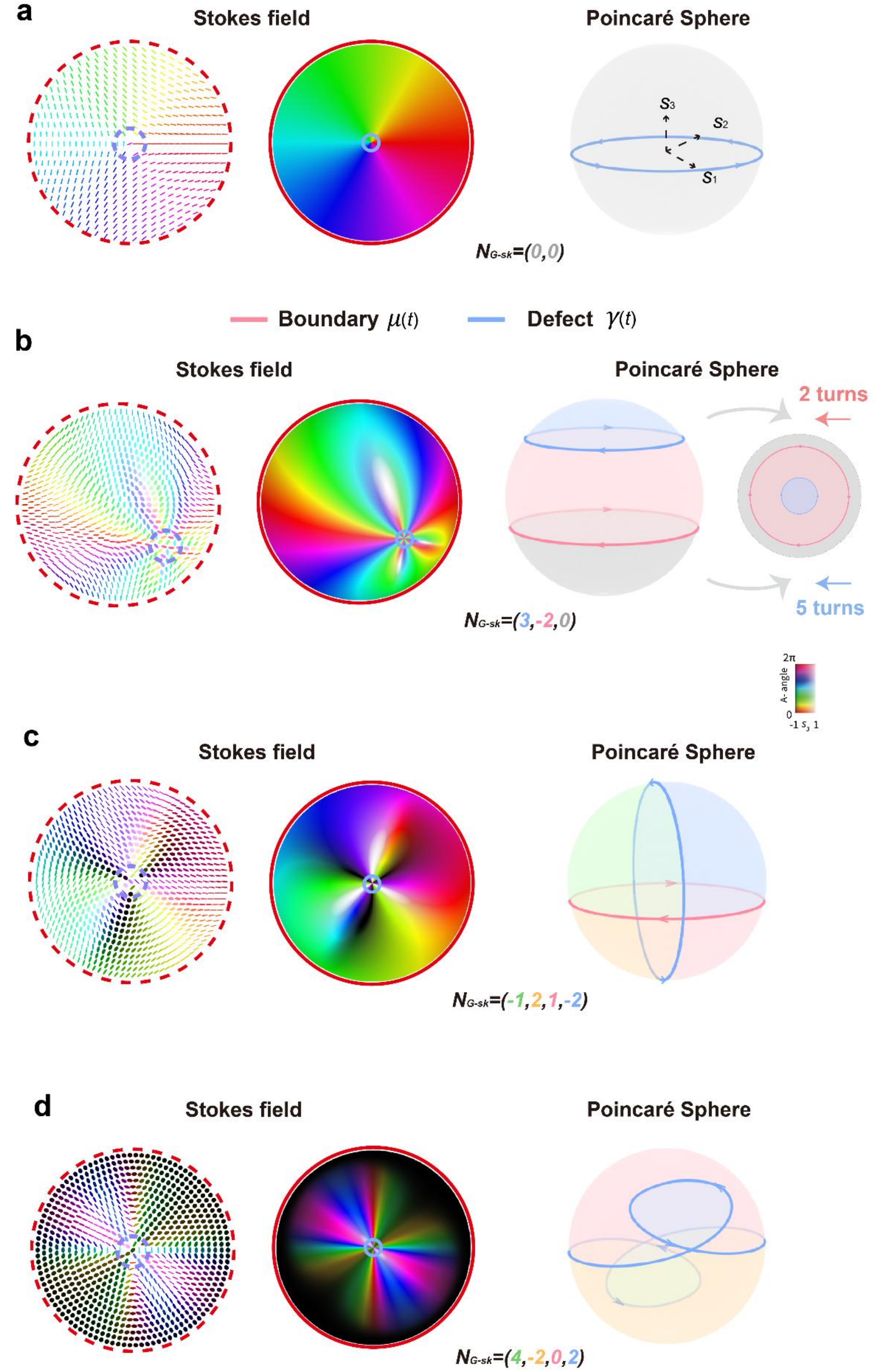

Fig. 5: **Boundary-trajectory configurations in different Stokes defect skyrmions.** Four representative configurations illustrate how the boundary image partitions the Poincaré sphere and determines the topological tuple. For each configuration, the left panels show the Stokes vector field and its corresponding color representation, as defined in Refs.[26,31], while the right panel shows the boundary trajectories and connected-region partition on the Poincaré sphere. The red line denotes the trajectory $\mu(t)$ generated by the boundary, while the blue line denotes the trajectory $\gamma(t)$ generated by the defect. **a.** An L-line-type configuration in which the Stokes field satisfies $s_3 = 0$ everywhere. Both the boundary image and the defect-induced trajectory map to the equator of the Poincaré sphere and coincide, so that the Stokes field does not enter either connected region, yielding $N_{G_sk} = (0,0)$. **b.** A non-constant boundary maps to a closed trajectory $\mu(t)$ on the Poincaré sphere. Together with the defect trajectory $\gamma(t)$, these two curves partition $S^2$ into three connected regions, yielding $N_{G_sk} = (3,-2,0)$. Here a projection of the spherical partition is shown for clarity. **c.** Intersections between the boundary $\mu(t)$ and $\gamma(t)$ further refine the partition of $S^2$ by introducing additional connected regions without requiring additional physical defects. In the configuration shown, the two trajectories intersect at two points, subdividing $S^2$ into four regions and yielding $N_{G_sk} = (-1,2,1,-2)$. **d.** A self-intersection trajectory $\gamma(t)$ partitions the Poincaré sphere into four connected regions using only one physical defect, producing the topological tuple $N_{G_sk} = (4, -2, 0, 2)$. Note that in some Poincaré-sphere visualizations, we rotate the coordinate system to make the partitioned regions clearer to the reader. The numerical values and trajectory plotting are strictly determined according to the theoretical framework.

**Different information densities and costs associated with different boundaries**

In this section, we show how changes in the boundary-trajectory geometry affect the dimension of the topological tuple and, consequently, the information density (Fig. 6). Here, the information density refers to the state count encoded across the components of the topological tuple within the same metasurface size and is therefore determined by the tuple dimension. In turn, the dimension is set by the number $R$ (represents the connected regions) formed by its boundary trajectories on the Poincaré sphere. A partition into $R$ regions produces a topological tuple in $\mathbb{Z}^R$.

Thus, as the simplest case (Fig. 6a), for mutually non-intersecting trajectories, each additional trajectory creates one further connected region, giving $R = N + 1$ ($N$ is the number of trajectories introduced by the defects). The corresponding topological tuple space in this case can reach $\mathbb{Z}^{N+1}$. A faster scaling is obtained when distinct trajectories intersect (Fig. 6b). For a generic arrangement in which every pair of trajectories intersects at two points and no three trajectories meet at the same point, the number of connected regions is $R = N^2 - N + 2$, giving quadratic growth in the dimension (tuple space can reach $\mathbb{Z}^{N^2-N+2}$; see Supplementary Note 1 for derivation details). Furthermore, a single trajectory can also intersect itself multiple times (see Supplementary Note 1). In this case, one physical defect can create multiple connected regions, so the tuple dimension can keep increasing as the trajectory becomes more complex (Fig. 6c).

We note that other designs may produce different information densities, but these should be chosen according to the requirements of specific applications, particularly the practical capabilities for information encoding and decoding. Overall, this framework reveals an unexplored nonlinear relationship between the number of defects and the achievable information density, paving the way for next-generation topological information encoding techniques.

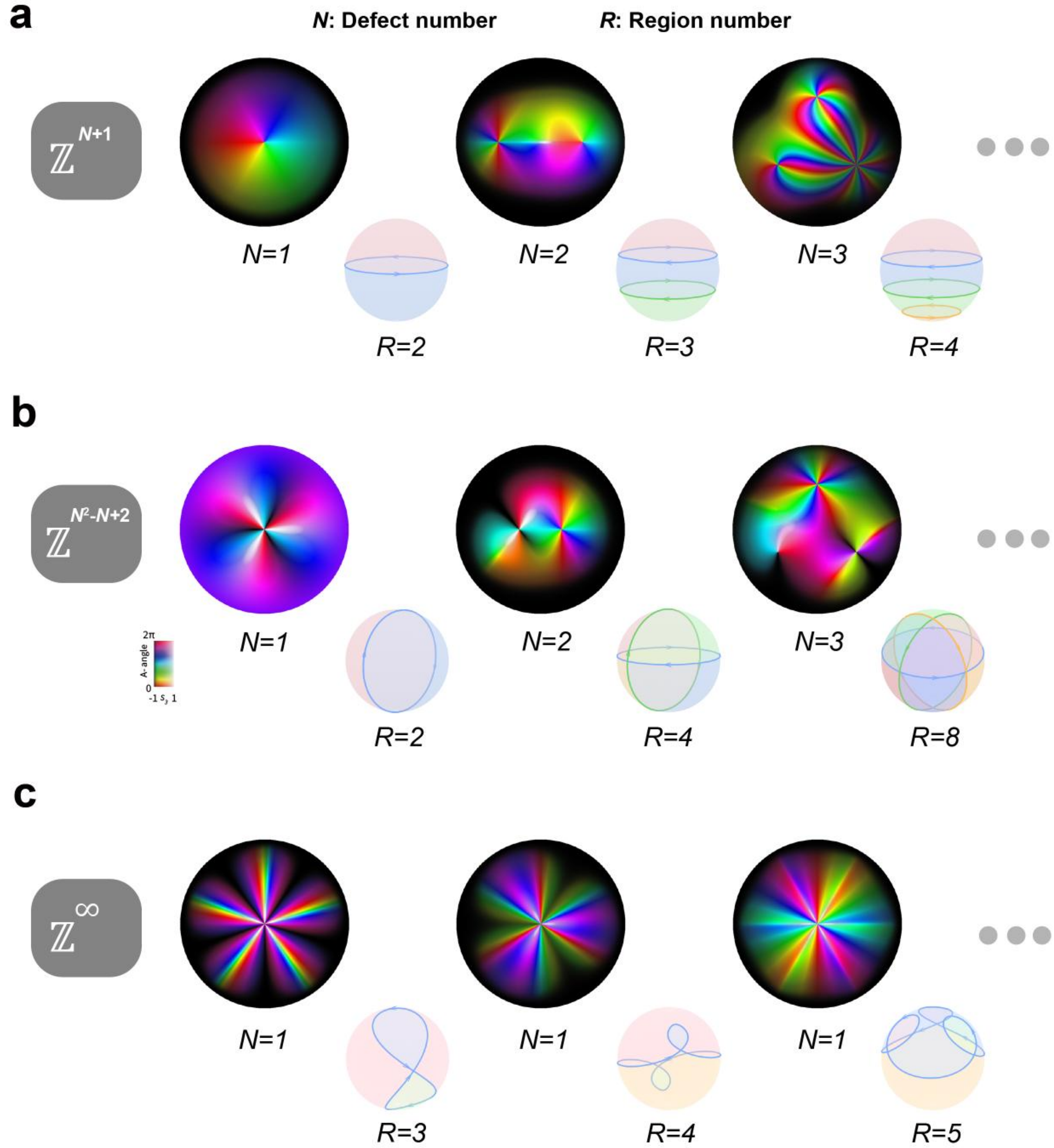

Fig. 6: **Topological information density under different trajectory geometries.** In each panel, the upper row shows the Stokes defect skyrmionic field, and the lower row shows the corresponding partition of the Poincaré sphere by the boundary image. $N$ denotes the number of defects and $R$ the number of connected regions, corresponding to the tuple dimension and thereby determining the information density. **a.** Mutually non-intersecting defect trajectories: each additional defect contributes one further latitudinal circle, so the region count grows linearly, $R = N + 1$ (here $N = 1,2,3$ give $R = 2,3,4$). Correspondingly, the tuple space is $\mathbb{Z}^{N+1}$ **b.** Pairwise-intersecting defect trajectories: every pair of trajectories crosses at two points, and the region count grows quadratically, $R = N^2 - N + 2$ (here $N = 1,2,3$ give $R = 2,4,8$). The tuple therefore lies in $\mathbb{Z}^{N^2-N+2}$ **c.** Self-intersecting trajectory of a single defect ($N = 1$): increasingly intricate self-crossings, conveniently described by spherical Lissajous curves, carve out arbitrarily many regions ($R = 3,4,5$ shown), so that $R$ is in principle unbounded even for a single defect, approaching $\mathbb{Z}^{\infty}$ in the dimension limit. Note that in some Poincaré-sphere visualizations, we rotate the coordinate system to make the partitioned regions clearer to the reader. The numerical values and trajectory plotting are strictly determined according to the theoretical framework.

**Acknowledgements**

We would like to acknowledge support from the Department of Engineering Science, University of Oxford and the Royal Society University Research Fellowship (URF\R1\241734) (C.H.), and the funding support from National Key R&D Program of China (2024YFA1209301), National Natural Science Foundation of China (NSFC) (62325503, 62422506, 12474383), Guangdong Provincial Quantum Science Strategic Initiative (GDZX2506004 (X.L.), GDZX2406004 (Z.L.D)). The authors thank Prof. Martin J. Booth (University of Oxford), Prof. Steve Morris (University of Oxford), Prof. Patrick Salter (University of Oxford), and An Aloysius Wang (University of Oxford) for valuable discussions and support.

**Author contributions**

Y.Z., F.L., R.Z., and C.H. conceived the main ideas, developed the concepts, and carried out the numerical simulations. Y.Z. and R.Z. performed the experiments and analyzed the experimental data. Y.Z., F.L., and Z.L.D. designed and fabricated the metasurfaces used in the experiments. Y.Z., F.L., R.Z., Z.L.D., Y.Z.X., Z.X.W., Y.X.Z., H.H.H., J.Z.Y., and C.H. prepared the figures and contributed to the interpretation and presentation of the results. Y.Z., F.L., R.Z., Z.L.D., and C.H. wrote and revised the manuscript. Z.L.D., X.P.L., H.H.H., J.Z.Y., and C.H. provided mentorship. X.P.L. and C.H. supervised the overall project. All authors reviewed the results, participated in discussions, commented on the manuscript, and approved the final version.

**Data availability**

The data that support the plots within this paper and other findings of this study are available from corresponding authors upon reasonable request.

**Conflict of interest**

The authors declare no competing interests.

**Additional information**

Correspondence and request for materials should be addressed to X.P.L. and C.H.